\documentclass[12pt,preprint]{aastex}

\slugcomment{Not to appear in Nonlearned J., 45.}

\shorttitle{NIR Polarimetry of S106}
\shortauthors{Saito et al.}

\begin{document}


\title{NEAR-INFRARED IMAGING POLARIMETRY OF S106 CLUSTER-FORMING REGION
       WITH SIRPOL}


\author{Hiro Saito\altaffilmark{1}, Motohide Tamura\altaffilmark{1,2},
Ryo Kandori\altaffilmark{1}, Nobuhiko Kusakabe\altaffilmark{1},\\
Jun Hashimoto\altaffilmark{2}, Yasushi Nakajima\altaffilmark{1}, 
Yaeko Sato\altaffilmark{2}, Tetsuya Nagata\altaffilmark{3},\\
Takahiro Nagayama\altaffilmark{3}, and Daisuke Kato\altaffilmark{4}}
\email{saito@nro.nao.ac.jp}





\altaffiltext{1}{National Astronomical Observatory of Japan, Osawa 2-21-1, 
Mitaka, Tokyo 181-8588, Japan}
\altaffiltext{2}{The Graduate University for Advanced Studies (SOKENDAI), Osawa 2-21-1, 
Mitaka, Tokyo 181-8588, Japan}
\altaffiltext{3}{Department of Astronomy, The University of Kyoto, Kitashirakawaoiwake, Sakyo-ku, Kyoto 606-8502, Japan}
\altaffiltext{4}{Department of Astronomy, School of Science, The University of Tokyo, Hongo 7-3-1,
Bunkyo-ku, Tokyo 113-0033, Japan}


\begin{abstract}
We present the results of wide-field $JHK_S$ polarimetry toward 
the H\,{\footnotesize II} region S106 using the IRSF (Infrared Survey Facility) telescope. 
Our polarimetry data revealed an extended (up to $\sim 5\arcmin$) polarized 
nebula over S106. 
We confirmed the position of the illuminating source of most of the nebula as consistent with 
S106 IRS\,4 through an analysis of polarization vectors. The bright portion of the polarized 
intensity is consistent with the red wing component of the molecular gas. 
Diffuse polarized intensity emission is distributed along the north--south molecular gas lanes.
We found the interaction region between the radiation from S106 IRS\,4 and the dense gas.
In addition, we also discovered two small polarization nebulae,  SIRN\,1 and SIRN\,2,
associated with a young stellar objects (YSO).
Aperture polarimetry of point-like sources in this region was carried out for the first time. 
The regional magnetic field structures were derived using point-like source aperture polarimetry, 
and the magnetic field structure position angle around the cluster region in S106 was found to 
be $\sim 120\arcdeg$.
The magnetic fields in the cluster region, however, have three type position angles: 
$\sim 20\arcdeg$, $\sim 80\arcdeg$, and $\sim 120\arcdeg$. 
The present magnetic field structures are consistent with results obtained by submillimeter 
continuum observations. We found that the magnetic field direction in the dense gas region is not consistent with that of the low density gas region.

\end{abstract}





\section{Introduction}

Sharpless 106 (S106, Sh-2-106, or G\,76.4-0.6) in Cygnus is a massive star-forming 
region associated with a bright bipolar nebula and a H\,{\footnotesize II} region 
(Sharpless 1959).
The first near-infrared observations of S106 were reported by Sibille et al.\,(1975), 
who found a near-infrared source and suggested that it could be the exciting star of 
the H\,{\footnotesize II} region S106. 
A bright infrared source, IRS\,4 (Gehrz et al. 1982), rests at the center of 
the bipolar structure, also designated as S106 IR (Sibille et al. 1982), 
IRS\,3 (Pipher et al. 1976), or PS (Felli et al. 1984).
The spectral type and luminosity of S106 IRS\,4 are estimated to be 
O8.5--O7 and $\sim 2$--$5 \times 10^{4} L_{\sun}$, respectively (Eiroa et al. 1979; 
Harvey et al. 1982; Mozurkewich et al. 1986; Martins et al. 2005).
Submillimeter continuum observations revealed a bright FIR source, referred to as 
S106 FIR, located $\sim 15\arcsec$ to the west of S106 IRS\,4 (Richer et al. 1993).
S106 FIR has been explained as a heavily embedded, luminous class 0 young stellar object (YSO), 
as this source was not detected at 20 $\micron$ (Gehrz et al. 1982; Richer et al. 1993).
In addition, Hodapp \& Rayner (1991) detected a cluster of 160 stars within the central 
0.3 pc radius of $K$-band ($\lesssim 14$ mag) imaging observations. 
Oasa et al.\,(2006) detected 600 embedded YSO candidates with near-infrared excesses 
within $\sim 5\arcmin \times 5\arcmin$ using the Subaru Telescope $JHK'$ survey.


The distance to S106 has been estimated as 0.5 -- 1.8 kpc by various methods.
The distance estimated by a photometric study based on $UBVRI$ photometry of 
field stars is $\sim 600$ pc (Staude et al. 1982).
However, that study essentially measured the distance to the first dust cloud in 
the direction of S106, not necessarily the distance to the object itself.
Recently, comparison between MSX mid-infrared images and CO molecular line data 
(Schneider et al. 2007) determined that the molecular cloud associated with S106 is 
a part of the Cygnus X molecular clouds and directly influenced by the UV radiation from 
Cyg OB1.
The distance of this cluster has been estimated to be 1.25 -- 1.83 kpc 
(Uyaniker et al. 2001), implying that S106 lies at the same distance.
They suggested that the obscuration leading to the closer photometric distance estimate
is due to a tenuous foreground molecular cloud unrelated to S106.
We use the rough average value of the obtained distances of $\sim 1.2$ kpc
as the distance to S106.


The molecular cloud in which S106 is embedded was found by Lucas et al.\,(1978).
The total mass of the cloud was estimated to be $\sim 7000$ -- $10000 M_{\sun}$ at 
a distance of 1.7 kpc (Bally \& Scoville 1982; Schneider et al. 2007).
The cloud is rotating around an axis at a position angle of $30\arcdeg$, i.e., with
the rotation axis in roughly the same orientation as the polar axis of 
the S106 bipolar nebula.
In addition, the southern lobe of S106 is brighter than the northern lobe
in the optical, although their radio fluxes are comparable (Bally et al. 1983;
Felli et al. 1984). This suggests that the northern lobe is more deeply embedded in 
the cloud and that the axis of the lobes is inclined with the northern lobe
away from the observer.


The magnetic field in S106 was studied with optical polarimetry and linearly polarized 
thermal emission from aligned dust grains.
Aperture polarimetry of 12 stars inside or near the bright nebula S106
within $6\arcmin \times 4\arcmin$ field of view was conducted by 
Hodapp \& Rayner (1991) using $I$-band images. The polarization
degree and polarization angle are estimated to be 0.65 -- 11.8\% and 
20 -- $94\arcdeg$, respectively.
In addition, Vall\'ee \& Fiege (2005) studied the magnetic field within
or near the bright nebula S106 based on the 850 $\micron$ continuum observations. 
They revealed that the magnetic field in the warm dense gas of the bipolar nebula 
away from the central dust lane appears roughly parallel along the bipolar nebula 
polar axis, but the magnetic field in the central dust lane appears horizontal and 
elongated along the lane. Note that the fields of view in these maps were too 
small ($\lesssim 3\arcmin \times 3\arcmin$) to reveal the magnetic field structure across 
the entire S106 cloud.
Polarized emission in S106 was studied with K-band polarimetry. 
Although Aspin et al.\,(1990) and McLean et al.\,(1987) revealed that the polarization vector
pattern is centrosymmetric about S106 IRS\,4,
the result obtained by Aspin et al.\,(1990) has a small field of view, though a high resolution 
($30\arcsec \times 30\arcsec$); the result obtained by McLean et al.\,(1987) has a large field of 
view ($5\arcmin \times 4\arcmin$) but a low resolution ($\sim 20\arcsec$).

In this paper, we present polarization images of the H\,{\footnotesize II} 
region S106, part of our ongoing project of $JHK_S$ polarimetry of star-forming regions.
Our observations were sufficiently deep ($J$ = $19.2$ mag at S/N = 10) and wide
 ($7\farcm7 \times 7\farcm7$), and of sufficiently high resolution ($\sim 1\farcs7$),
covering a large extent of S106.
Our wide-field polarization images can reveal large-scale infrared reflection
nebulae (IRNe) and delineate the magnetic field structure through measurements
of point source polarizations.





\section{Observations and Data Reduction}

The observations of S106 were carried out on 2006 June 15 with the imaging 
polarimeter SIRPOL, the polarimetry mode of the SIRIUS camera 
(Simultaneous-3color Infrared Imager Unbiase Survey : Nagashima et al. 1999;
Nagayama et al. 2003) for Kandori et al.\,(2006), on the 1.4 m telescope IRSF 
at the South Africa Astronomical Observatory (SAAO).
 SIRPOL is composed of an achromatic (1--2.5 $\micron$) wave plate rotator and 
a high extinction ratio wiregrid analyzer unit located on the upstream side of 
the $JHK_S$-simultaneous imaging camera SIRIUS.
The imaging scale of the array is $0\farcs45$ pixel$^{-1}$, giving 
a field of view of $7\farcm7 \times 7\farcm7$. 

The polarization was measured by stepping the half-wave plate to four angular 
positions ($0\arcdeg$, $22\fdg5$, $45\arcdeg$, $67\fdg5$).
We made 10-s exposures per wave plate position at 10 dithered positions (1 set),
and we repeated the same set four times for each object. Sky frames were taken 
using the same method. The total integrated time was 400 s per wave plate angle.
The typical seeing during the observations was $\sim 1\farcs7$ (3.7 pixels) at $J$.

We reduced the observed data using standard infrared image reduction method
(flat-field correction, median sky subtraction, and frame registration).
Stokes $I, Q, U$ parameters were obtained using 
$I = (I_{\rm 0\arcdeg} + I_{\rm 22\fdg5} + I_{\rm 45\arcdeg} + I_{\rm 67\fdg5})$/2, 
$Q = I_{\rm 0\arcdeg} - I_{\rm 45\arcdeg}$, 
and
$U = I_{\rm 22\fdg5} - I_{\rm 67\fdg5}$.
The degree of polarization, $P$, and the polarization angle, $\theta$, were
estimated using $P = \sqrt{Q^{2} + U^{2}}/I$ and 
$\theta =  0.5 \times tan^{-1}(U/Q)$, respectively.
Since the polarization degree, $P$, is a positive quantity, the polarization
degree would be overestimated by the error in $P$. Therefore, we calculated
the debiased polarization degree, $P_{\rm db}$, to remove the effect of
the error using $P_{\rm db} = \sqrt{P^{2}-\Delta P^{2}}$, where $\Delta P$ is
the error in $P$ (Wardle \& Kronberg 1974). Hereafter, we regard the debiased
polarization degree as the polarization degree.
We calibrated the $I$ sky level using 2MASS\footnote{The Two Micron All Sky 
Survey (2MASS) is a joint project of the University of Massachusetts and 
the Infrared Processing and Analysis Center/California Institute of Technology, 
funded by the National Aeronautics and Space Administration and the National 
Science Foundation (http://pegasus.phast.umass.edu).} images covering a much 
larger field of view.
The 10 $\sigma$ limiting magnitudes for the surface brightness of $I$ were
19.2, 18.9, and 18.0 mag arcsec$^{-2}$ for $J$, $H$, and $K_S$, 
respectively.

For source detection and photometry on the Stokes I image, we used the 
IRAF\footnote{IRAF is distributed by the US National Optical Astronomy 
Observatories, which are operated by the Association of Universities of 
Research in Astronomy, Inc., under cooperative agreement with the National
Science Foundation.} DAOPHOT package (Stetson 1987).
We detected stars with a peak intensity greater than 5 $\sigma$ above
the local background and measured the instrumental magnitudes of the stars 
using aperture photometry with an aperture radius of 3.7 pixels. The number of 
stars detected with a photometric error of less than 
0.1 mag were 627, 1228, and 1376, respectively. In addition, 583 stars were detected in 
all three bands. The limiting magnitudes in this case were 18.0, 17.5, and 16.5 mag, respectively.
Next, we performed a photometric calibration using the 2MASS point source catalog.
The magnitude and color of our photometry were transformed into the 2MASS 
system using
\begin{eqnarray}
{\rm MAG_{\,2MASS}} = {\rm MAG_{\,IRSF}} +\alpha_{\rm 1} \times {\rm COLOR_{\,IRSF}}
+ \beta_{\rm 1},
\end{eqnarray}
\begin{eqnarray}
{\rm COLOR_{\,2MASS}} = \alpha_{\rm 2} \times {\rm COLOR_{\,IRSF}} +\beta_{\rm 2},
\end{eqnarray}
where the coefficients $\alpha_{\rm 1}$ are $0.057$, $-0.013$, and $-0.046$ for 
$J$, $H$, and $K_S$, respectively. The coefficients $\alpha_{\rm 2}$ are 
$1.070$ and $1.019$ for $J - H$ and $H - K_S$, respectively.
Note that the coefficients $\beta_{\rm 1}$ and $\beta_{\rm 2}$ include both
the zero point and aperture correction.

\section{Results}

\subsection{Large-Scale Intensity and Polarized Properties in S106}
Although a high-resolution ($0\farcs35$) $JHK'$ intensity image of S106 taken
with Subaru was reported previously (Oasa et al.\ 2006), no wide-field polarimetry 
observations toward S106 have been performed. Our observations with SIRPOL revealed 
the near-infrared polarization distribution in S106 for the first time.
Figures 1a and 1b show the $JHK_S$ color-composite intensity 
(Stokes $I$) image and the $JHK_S$ color-composite polarized intensity
($PI$) image, respectively.

In Figure 1a, the diffuse emission of the northern and southern lobes are 
extended more than the field of view of the Subaru images, and Figure 1b shows 
the first characteristic polarization nebulosity over the S106 region. 
Ridges of polarized intensity exist on the eastern and western edges of nebula S106.
In addition, the southern edge of the southern lobe of S106 has a relatively strong 
polarized intensity.
The northern and southern parts of the strong polarized intensity around 	
the exciting source of S106, S106 IRS\,4, have fan-shaped and arc-edge 
structures, respectively.
In addition, a dark spot $8\arcsec$ in size, is located $15\arcsec$ to 
the northeast of S106 IRS\,4.

Next, we present a polarization vector map in Figure 2 to identify the location 
of the illuminating source(s) of the infrared reflection nebula.
We found that the vector patterns of the high-intensity part of the nebula
appear clearly centrosymmetric and the perpendicular lines of each 
vector point to a position at the exciting star of S106, S106 IRS\,4, 
although the vector patterns of the low-intensity part 
(north and southeast part of the nebula) fall into disorder.
This indicates that the exciting star, S106 IRS\,4, is the illuminating star of the nebula.

\subsection{Small-Scale Infrared Reflection Nebulae in S106}
We found two small-scale infrared reflection nebulae (SIRNe\,1 and 2), 
shown in Figure 1b. One of them, SIRN\,1, is associated with a bright star 
classified into PMS type using our $JHK_S$ bands data (see Section 3.3). 
Although the other nebula, SIRN\,2, has no clear point sources 
in our observations, this nebula is associated with one class I like object,
No.\,258, which was identified by Oasa et al.\,(2006).

$PI$ and the vector maps of SIRN\,1 are shown in Figure 3a. Note that these 
images are formed by subtracting the extended component of the large reflection 
nebula formed by S106 IRS\,4 from the original images.
The pattern of the polarization vector clearly appears centrosymmetric and 
the perpendicular lines of each vector roughly point to the central star. 
These indicate that the nebulosity is illuminated by the central star. 
In addition, the $PI$ image of SIRN is clearly butterfly-shaped, with 
the dark lane extending perpendicular to the bright region.
This characteristic agrees with the polarization picture of the disk-envelope
system around young stars (e.g., Nagata et al. 1983).

$PI$ and the vector maps of SIRN\,2 are shown in Figure 3b.
The SIRN exhibits an extended emission pattern and the polarization vector 
is perpendicular to the elongated direction of the emission pattern.
This feature agrees with the polarization picture of a monopolar outflow lobe. 
From the results of Oasa et al.\,(2006), the faint source No.\,258,
classified as a class I like object, is associated with
the north edge of the nebula, and the perpendicular lines of each vector
roughly point to source No.\,258. From this result, we suggest that
this reflection nebula is a bipolar nebula driven by the class I like
object, source No.\,258.

\subsection{Infrared Colors of Stars}
The NIR photometry of the stars provides information on large embedded 
populations in the cloud. Figure 4 shows the $J - H$ versus $H - K_S$ 
color-color diagram for the identified sources. We employ the reddening law
of $E(J - H)/E(H - K_S)$ = 1.7 (Koornneef 1983).
This diagram provides a good tool for discriminate between the interstellar reddening
and intrinsic color excess.
We used this diagram for all sources detected in the $JHK_S$ bands to 
classify sources into two groups (MS star/dwarf+giant and PMS star/protostar). 
We found 69 stars in the PMS star/protostar regions (filled circles)
and 514 stars in the MS stars/dwarf+giant region (open circles).
In addition, the stars in the MS stars/dwarf+giant region are divided into
two groups, stars with high reddening and those with no reddening. 
The stars with no reddening are the foreground stars (FG stars). 
Thus, we found that the 363 stars with high reddening are the stars in 
the S106 region and background stars (S106 stars).

Next, we checked the surface stellar density of S106 stars to reveal the effect
of the background stars. Figure 5 shows the distribution of the surface stellar 
density of S106 stars superposed on the distribution of the $^{13}$CO integrated
intensity obtained by Schneider et al.\,(2007).
The distribution of the surface stellar density dramatically decreases with 
distance from S106 IRS\,4 and reaches a minimum near $\sim 2\farcm2$. 
In the region at $>2\farcm2$, the surface stellar density gently increases with distance, 
particularly for the north region.
This feature suggests that the number of background stars increases with distance 
from S106 IRS\,4 in the north region.
The distribution of the molecular gas in S106 region has 
a sharp edge north of S106, and in the north region, we see an anti-correlation 
between the surface stellar density and the molecular gas distribution. 
Therefore, we could have detected many background stars in the north region because 
the H$_2$ column density dramatically decreases with distance from S106 IRS\,4.
This indicates that stars at distances of $>2\farcm2$ from S106 IRS\,4 are 
mixed with many background stars.
Thus, we regard the stars at $<2\farcm2$ from S106 IRS\,4 as members of the cluster in 
S106 (CL stars) and regard the other stars at $>2\farcm2$ as both stars around S106 and 
background stars (field stars). 

\subsection{Aperture Polarimetry}
We need to perform interstellar polarimetry to determine the magnetic
field structure on the parsec scale. Therefore, we measured software 
aperture polarizations of point-like sources detected in the field of view.
We rejected the sources with photometric errors greater than 0.1 mag and 
the polarization degree to error ratios smaller than 2.
In addition, we rejected bright saturated sources. Note that the position
angle error of each vector is less than $15\arcdeg$.

We measured the polarization degree and polarization angle of 162 source
(43 FG stars, 55 CL stars, and 64 field stars).
These sources include 8 of 12 sources with estimated polarization degree and
angle from Hodapp \& Rayner (1991). The polarization angles obtained by 
the present study were roughly consistent with those previous values.
We plotted these sources in the $P_{\rm H}$ versus $H - K_S$ diagram shown
in Figure 6. A linear fit to the relationship is approximately obtained as
$P_{\rm H}$ = $4.4 \pm 0.3 \times (H - K_S)$, except for the YSO candidates.
This is smaller than the slopes ($\sim 6.0$) of the relationships in the M42 and NGC 2071 
regions (Tamura et al. 2007; Kusakabe et al. 2008). 
The observational upper limit of the relationship was obtained by Jones (1989), and 
we represent this upper limit with a dashed line in Figure 6.
In addition, we see no clear difference in index between the relationships 
for CL stars ($\sim 4.0$) and field stars ($\sim 4.7$).
Therefore, although the polarization efficiencies have similar values in 
the whole molecular clump, little difference exists in the polarization 
efficiency among the star-forming regions.

\section{Discussion}

\subsection{Comparison of the Polarized Intensity and Molecular Gas Distribution}
To reveal an interaction between the radiation from the massive star S106 IRS\,4 
and the molecular gas, we compared the infrared polarized emission and CO emission
distributions. Schneider et al.\,(2007) revealed the full extent of the S106 molecular cloud, 
which has a mass of 7600 $M_{\sun}$ at a distance of 1.7 kpc using 
the $^{13}$CO ($J\!=\!1$--$0$) line. 
In addition, Schneider et al.\,(2002) revealed the structure of the dense gas around 
S106 with a high resolution of $11\arcsec$ using CO ($J\!=\!2$--$1$) lines.
Figure 7 shows the $PI$ image superposed on the distributions of four velocity 
planes of the $^{13}$CO ($J\!=\!2$--$1$) obtained by Schneider et al.\,(2002). 
Figures 7a and 7d are the CO distributions with typical velocities for 
the blue wing and the red wing, respectively. Figures 7b and 7c are 
the velocity component corresponding to the dark lane in the infrared emission
and the central velocity of the S106 molecular cloud, respectively.

The blue wing emission of the outflow component in Figure 7a is compact 
and located at S106 IRS\,4. The red wing emission in Figure 7d extends father around 
S106 IRS\,4 with a size of $\sim 1\arcmin$.
In particular, the strong part of the red wing emission is located at 
the bright part of the $PI$ image.

Next, the dark lane component of the molecular cloud in Figure 7b corresponds 
to the upper side of the dark lanes in the polarized infrared emission.
The main component of the molecular cloud in Figure 7c shows an elongated 
north--south distribution along both sides of the near-infrared cavity walls. 
The diffuse emission in the $PI$ image is well correlated with the north--south lanes of 
the CO main component, and this feature indicates the possibility of interaction between 
radiation from S106 IRS\,4 and the dense molecular gas.
In particular, the polarized emission associated with the east side lane is 
bright and the emission boundary is entirely consistent with that of 
the east side lane. 
In contrast, the polarized emission of the west side lane CO distribution boundary is 
much weaker than that of the east side lane.
In the west of S106 IRS\,4, a dark spot in the polarized emission exists at offsets
($+5\arcsec$,$-4\arcsec$) from H$_2$O masers/S106 FIR (e.g., Furuya et al. 1999),
and a compact molecular core, identified by CS and $^{13}$CO lines 
(e.g., Barsony et al. 1989), is located at the dark spot. In addition, a large dark bubble 
in the (polarized) infrared emission rests on the west side of the dark spot. 

From these features, we show the relationship between the gas structure and 
polarization intensity in Figure 8.
First, a bright $PI$ in the central region of S106 appears probably due to the dense part of 
the outflow lobes existing on the far side of S106 but not near side.
Basically, to generate bright polarized emission, starlight must  be reflected by the material. 
In addition, for us to detect the bright polarized emission, the dense gas must not be located 
at the near side of the reflected area.
Next, the difference in polarized emission brightness between the east and west lanes
could be caused by the core near S106 IRS\,4 blocking radiation from S106 IRS\,4.
Therefore, we suggest that the polarized emission is 
weak because almost all of the radiation from S106 IRS\,4 fails to reach the front of 
the west side lane.

%

From these features, we suggest that the molecular gas in the west lane would not be remarkable 
affected by the radiation from the massive star. 
As to star formation activity, Motte et al.\,(2007) and Oasa et al.\,(2006) revealed the YSOs and dense cores 
distributions, and they found that the distributions have no clearly different between these lanes.
Thus, although the influence of the radiation from the massive star to the molecular gas would be 
clearly different between the east and west lanes, it is not clear difference in star formation activity 
between these lanes.
These results suggest that the influence of the radiation from the central star on star formation and 
core formation is not large.

\subsection{Magnetic Field Structures}
The aperture polarimetry of stars provides important information on 
the magnetic field structure. If the grains are aligned by magnetic fields,
we can infer the direction of the magnetic fields projected onto the sky from
the direction of the stellar polarization vectors (e.g., Weintraub et al. 2000).
Although a grain alignment mechanism caused by the streaming motion of the gas
around sources with molecular outflows may be the important process, 
our assumption is plausible because such grain alignment was observed in 
a molecular core without star formation (Kandori et al. 2008).

The histograms of the polarization position angles of the FG stars, CL stars, and 
field stars in S106 region are shown in Figure 9.
Although the position angles of both FG stars and CL stars are distributed randomly, 
the field star angles clearly peak at $\sim 120\arcdeg$.
The field stars record interstellar polarization by the S106 molecular cloud because 
the field stars include many background stars.
Thus, the large-scale distribution of the magnetic field in the S106 molecular
cloud is expected to be in the direction of $\sim 120\arcdeg$.
Note that although the magnetic field position angle of the surrounding the S106 molecular 
cloud was estimated to be $\sim 55\arcdeg$ by optical polarimetry
(Staude et al. 1982), all the stars which estimated the position angle by optical polarimetry
are included in the area of foreground stars in Section 3.3 according to 
the 2MASS point source catalog.

Although CL stars have to exist in the S106 molecular cloud, the distribution of 
CL star position angles has no clear peak. The main cause of this feature would be 
an absence of dense gas on the near side of the cluster, or the influence of molecular outflow.
Most of the dense gas on the near side of CL stars would be ionized by 
UV radiation from S106 IRS\,4 or scattered by molecular outflows, or else
the grain alignment would be disturbed by the molecular outflow.
Indeed, as the $^{13}$CO intensity of the optical lobe of S106 nebula is weaker
than the north-south walls (Schneider et al. 2002), most CL stars in 
the S106 nebula would not match the interstellar polarization.

The aperture polarization vector map of point-like sources in the $H$-band
superposed on the intensity image is shown in Figure 10.
Although the dispersion of the polarization vector directions is large,
the overall polarization angle distribution of the outside the cluster area 
(the area outside the dashed circle in Figure 10) is $\sim 100$ -- $150\arcdeg$. 
In addition, the polarization angle distribution in the west--northwest 
region of the S106 nebula is $\sim 80\arcdeg$.
The region with the $\sim 80\arcdeg$ polarization angle lies just east
of the $^{13}$CO component corresponding to the dark lane in Figure 7b.
The polarization angle in the cluster area (the area inside the dashed circle 
in Figure 10) is complicated. Although the angle in the area corresponding to 
the dark lane is $\sim 80\arcdeg$, most of the polarization angle inside 
or near the bright nebula is $\sim 20\arcdeg$ and roughly parallel to 
the bipolar nebula polar axis.

Vall\'ee \& Fiege (2005) studied the warm dense gas outside the hot gas through
the 850 $\micron$ continuum observations. They found that the magnetic field in 
the bipolar nebula away from the central dust lane, corresponding to the dark 
lane component, is roughly parallel to the polar axis of the bipolar nebula. 
This result is roughly consistent with our finding. 
The magnetic field in the central dust lane, however, appears horizontal and 
elongated along the lane. 
The dust lane detected by Vall\'ee \& Fiege (2005) is sandwiched between 
the center region of the bipolar nebula and the west--northwest region with 
the $\sim 80\arcdeg$ polarization angle.
Thus, the magnetic field in the dark lane region, corresponding to 
the warm dense gas region except for the region inside or near 
the nebula, has the same direction all over.
From these results, we found that the magnetic field direction of 
the high dense gas region of the S106 cloud disagrees with that of
the low dense gas region.
We thus conclude that dichroic polarization at near-infrared wavelengths is
a good tracer of magnetic fields in regions with wide density and scale range.

Finally, we discuss the relationship between the gas structure and the magnetic field 
structures in the S106 cloud. 
Figure 11 shows the relationship between the gas structure and the magnetic field structure. 
The head of the cloud exhibits a spherical structure with an axial rotation direction 
of $\sim 30\arcdeg$.
The direction of the magnetic field in the whole cloud is roughly vertical along
the rotation axis.
In contrast, the direction of the magnetic field in the dense gas (e.g., dust lane)
is roughly parallel to the dust lane with an offset of $\sim 40\arcdeg$ from 
the magnetic field of the whole cloud.
From these results, we suggest that such magnetic fields would be formed by contraction 
while the gas/dust coupled with a magnetic field rotates in the cloud.
For a detailed discussion on such a  magnetic field evolution, we must consider the relationship
between the magnetic field in various clouds with star formation and the gas structure
observed at high resolution, such as is obtainable with ALMA (Atacama Large 
Millimeter/submillimeter Array).

\section{Conclusion}
We conducted deep, wide-field $JHK_S$ imaging polarimetry toward
S106, a large bipolar nebula. This is the first imaging polarimetry covering
the entire bright bipolar nebula including a young cluster. The main results of 
the present study are summarized as follows:

1. We found a clear and extended infrared reflection nebula (IRN) over S106 on
    our polarization image. We confirmed that the illumination source of 
    the nebula is the central massive star, S106 IRS\,4 through an analysis of
    the polarization vectors.

2. We discovered two small-scale IRNe (SIRNe) associated with a YSO in 
   our polarization images. The pattern of the polarization vector around 
   the star appears to be centrosymmetric, which indicates that the SIRNe are 
   illuminated by the associated YSO.

3. We classified 583 point-like sources into three categories: foreground
   stars, members of the S106 cluster and stars around S106. In addition,
   we measured the polarization degree and polarization angle of 250 of the 
   point-like sources detected on the intensity images at $JHK_S$.
   
4. We found that the east boundary of the polarization image is consistent with
   the boundary of the east side lane of the molecular gas. 
   In addition, the west boundary of the polarization image is very weak, as 
   almost all of the radiation from S106 IRS\,4 fails to reach the front of the west 
   side lane because the core located at the dark spot near S106 IRS\,4 blocks 
   the radiation from S106 IRS\,4.

5. The magnetic fields of the area around the bright nebula S106 derived from 
   the dichroic polarization run at a position angle of $\sim 120\arcdeg$ projected 
   on the sky. The magnetic fields of the area near or inside the bright nebula S106 
   run at a position angle of $\sim 20\arcdeg$ projected on the sky.

We are grateful to the referee and Tomoyuki Kudo for their helpful comments and suggestions. 
We thank Satoshi Mayama for his valuable advice on the analysis.
We thank N. Schneider for kindly providing the $^{13}$CO data in FITS format.
This study made use of the SIMBAD database.
The IRSF/SIRIUS project was initiated and supported by Nagoya University, 
the National Astronomical Observatory of Japan, and the University of Tokyo in
collaboration with the South African Astronomical Observatory under a financial
support of Grants-in-Aid for Scientific Research on Priority Areas (A) 
No. 10147207 and No. 10147214, and Grants-in-Aid No. 13573001 and No. 16340061
of the Ministry of Education, Culture, Sports, Science and Technology, Japan.

\begin{figure}
\begin{center}
\epsscale{0.9}
\plotone{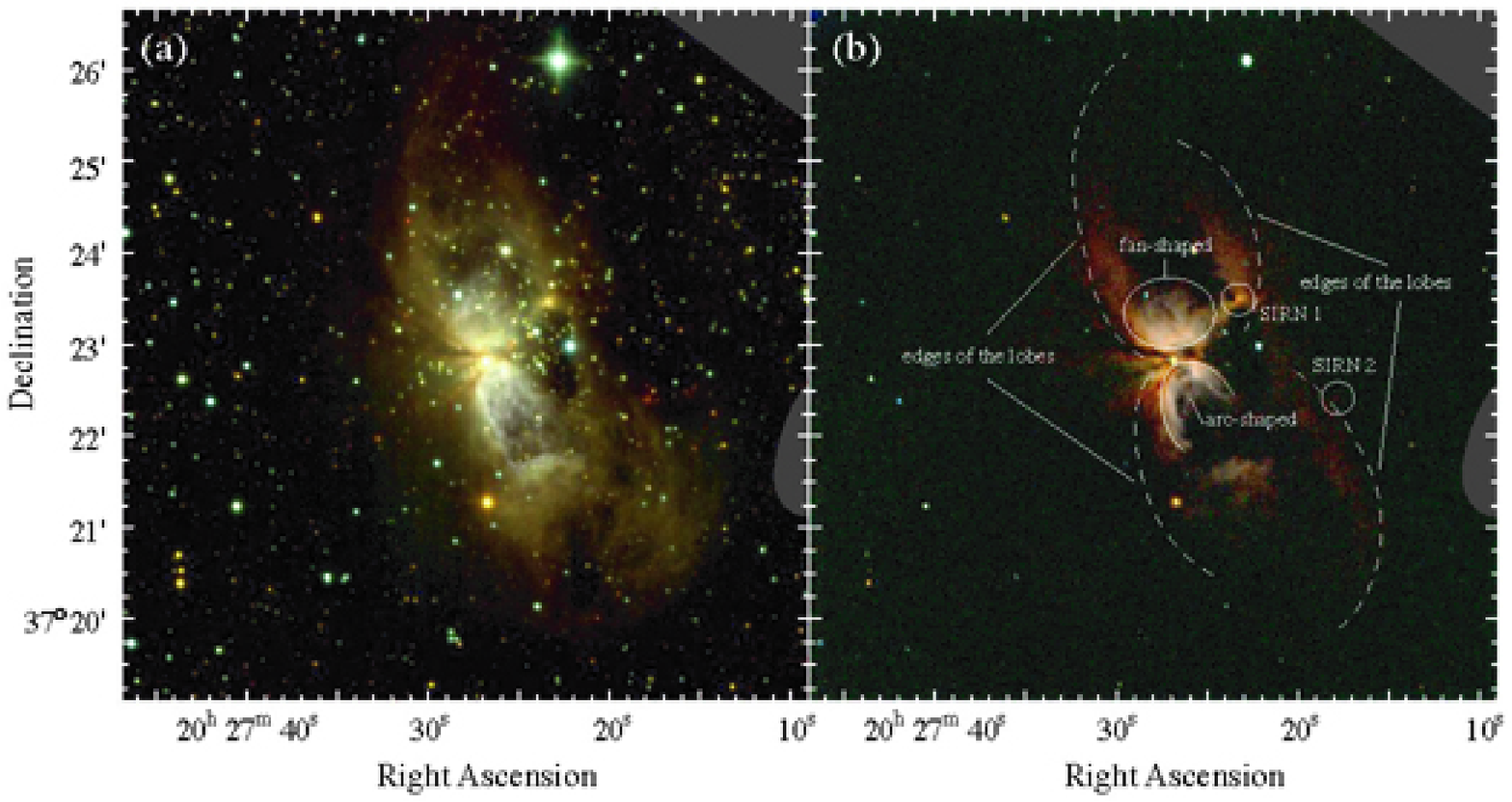}
\caption{
           (a) The $JHK_S$ composite images of the intensity image toward S106. 
           (b) The $JHK_S$ composite images of the polarized intensity image. 
           Two newly discovered small infrared nebulae are enclosed by circles.
           The dotted curves denote the edge of the lobes of the nebula. 
           The ellipse and the solid line indicate the fan-shaped and arc-shaped structure,
           respectively (discussed in Section 3.1).
           We note that the presence of bad pixel clusters on the $J$-band image 
           around the upper-right corner and the middle of right side, both masked in gray.
           }
\label{Figure 1}
\end{center}
\end{figure}


\begin{figure}
\begin{center}
\epsscale{0.70}
\plotone{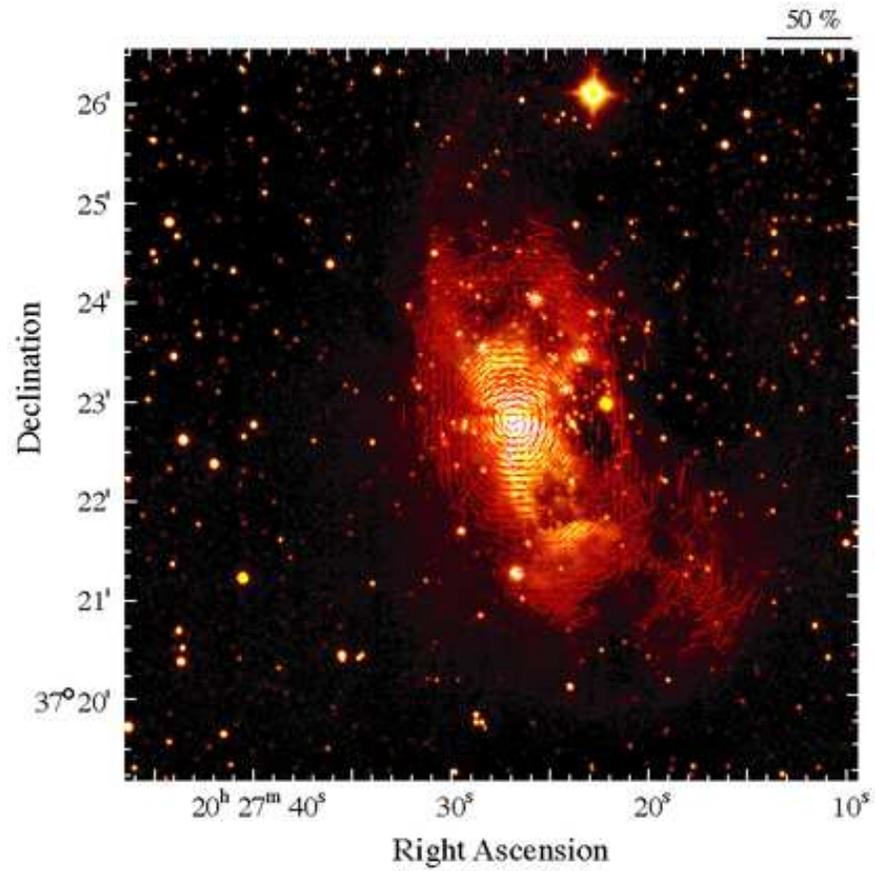}
\caption{
           Polarization vector map of the intensity image in the $H$-band 
           ($I/\Delta I > 10$). Polarization vectors for the low-intensity
           region ($I/\Delta I < 10$) are not plotted. 
           }
\label{Figure 2}
\end{center}
\end{figure}

\clearpage 

\begin{figure}
\begin{center}
\epsscale{0.55}
\plotone{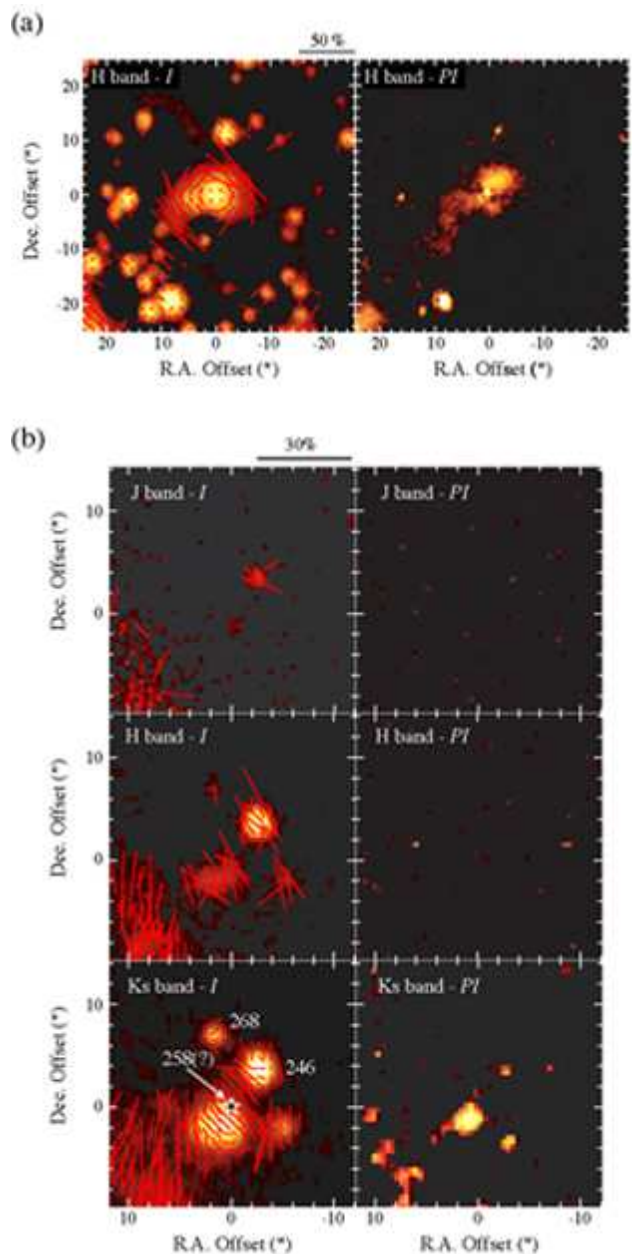}
\caption{
           Left column: Polarization vector maps of the intensity images.
           The polarization vectors of the low-intensity region
           ($I/\Delta I < 10$) are not plotted. Right column:Polarized 
           intensity images.  
           (a) The images formed by subtracting the extended component of
                 the large reflection nebula formed by S106 IRS\,4 from 
                 the original images. The intensity and polarized images are in the $H$-band.
                 The ($0\arcsec$,$0\arcsec$) coordinates are 
                 ($20^{\rm h} 27^{\rm m} 23\fs3$, $37\arcdeg 23\arcmin 26\fs0$). 
           (b) The labels are the number of YSO candidates identified by Oasa et al.\,(2006).
                 The ($0\arcsec$,$0\arcsec$) coordinates are 
                 ($20^{\rm h} 27^{\rm m} 17\fs8$, $37\arcdeg 22\arcmin 23\fs2$).
           }
\label{Figure 3}
\end{center}
\end{figure}

\clearpage 

\begin{figure}
\begin{center}
\epsscale{0.43}
\plotone{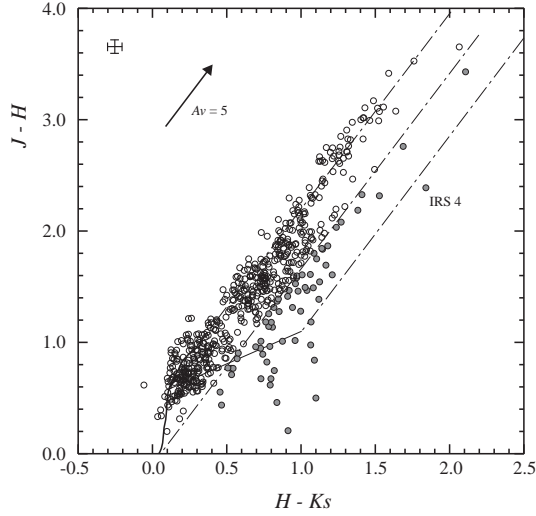}
\caption{
           Color-color diagram of all point-like sources. The bold curves represent 
           the sequences for field dwarfs and giants taken from 
           Bessell \& Brett (1988), and the thin line represents the locus of 
           T Tauri stars (Meyer et al. 1997). The dash-dotted lines are drawn 
           parallel to the reddening vector calculated using the reddening low
           from Koornneef (1983). Open and filled circles denote 
           MS stars/dwarf+giant stars and PMS stars/protostar candidates, 
           respectively.
           }
\label{Figure 4}
\end{center}
\end{figure}

\begin{figure}
\begin{center}
\epsscale{0.45}
\plotone{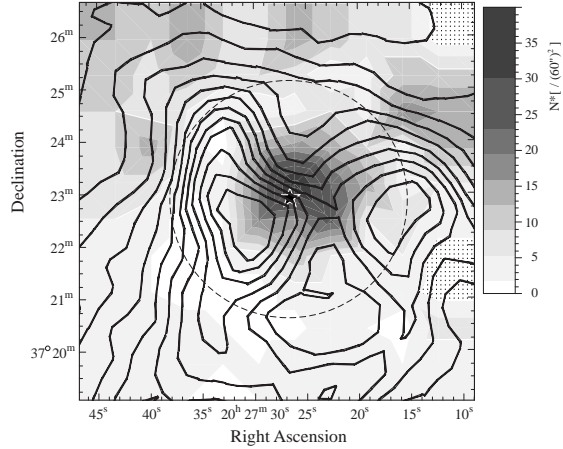}
\caption{
           The distribution map of the surface density of S106 stars, derived from  
           all stars except foreground stars. 
           The $^{13}$CO ($J\!=\!1$--$0$) integrated intensity contour map
           is superposed.
           The star and dotted circle denote the position of S106 IRS\,4 
           and the distance of $2\arcmin2$ from S104 IRS\,4, respectively.
           The regions masked by hatched dots denote the bad pixel clusters on the $J$-band image. 
           }
\label{Figure 5}
\end{center}
\end{figure}

\clearpage 

\begin{figure}
\begin{center}
\epsscale{0.5}
\plotone{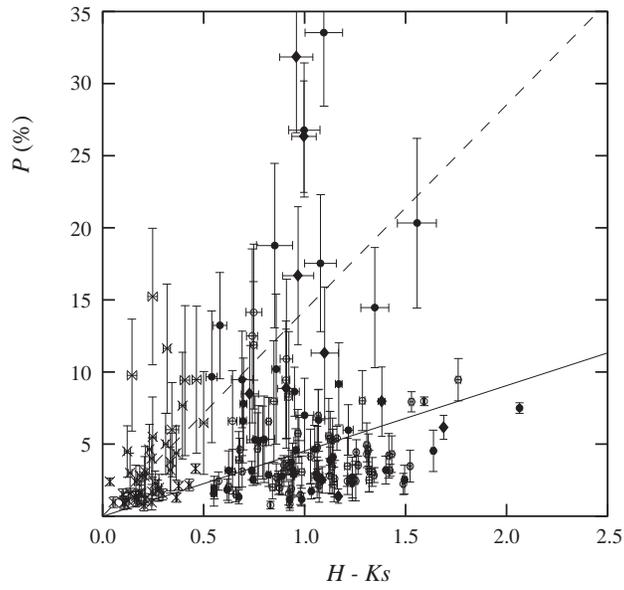}
\caption{
           Correlation between the polarization degree in the $H$-band and
           the $H - K_S$ color. The crosses, open circles, filled circles,
           and diamonds denote foreground stars, field stars, CL stars,
           and YSO candidates, respectively. The dashed line denotes
           the upper limit of the molecular cloud relationship (Jones 1989) 
           and the solid line is the result of linear fitting to all data 
           points except YSO candidates.
           }
\label{Figure 6}
\end{center}
\end{figure}

\clearpage 

\begin{figure}
\begin{center}
\epsscale{0.8}
\plotone{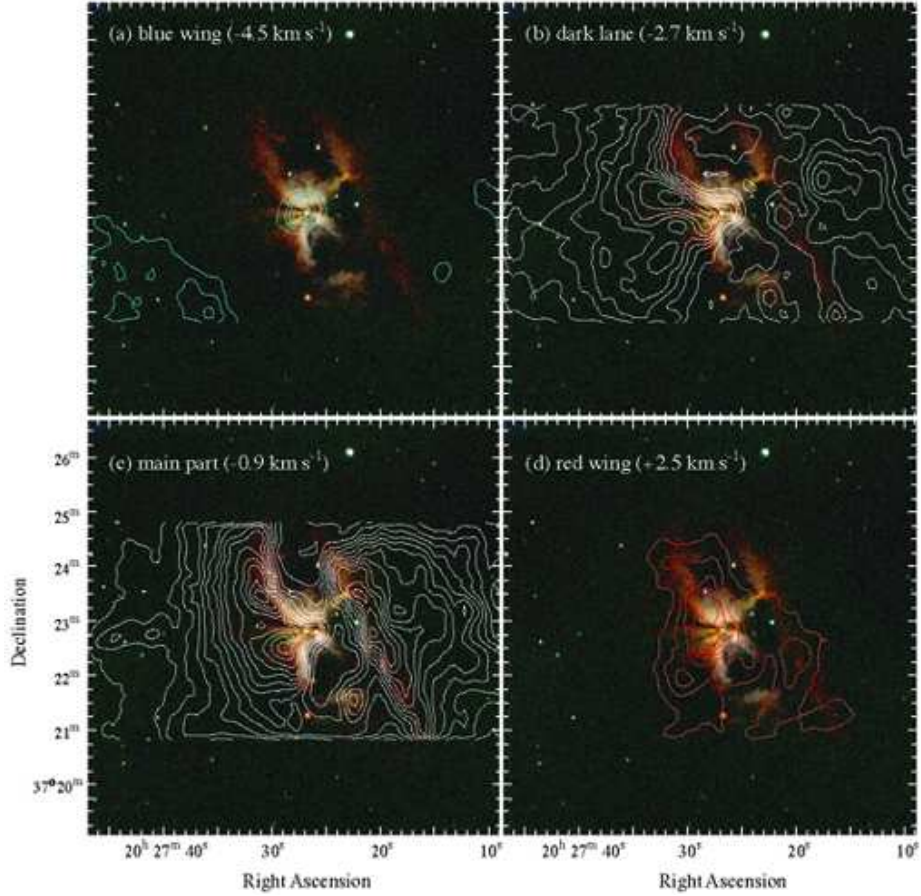}
\caption{
           Four different velocity panels of $^{13}$CO ($J\!=\!2$--$1$) 
           emission obtained by Schneider et al.\,(2002) overlaid on 
           the $JHK_S$ polarized intensity image.
           The contour levels are from 1.3 K\,km\,s$^{-1}$ in steps 1.3 K\,km\,s$^{-1}$.
           }
\label{Figure 7}
\end{center}
\end{figure}

\begin{figure}
\begin{center}
\epsscale{0.5}
\plotone{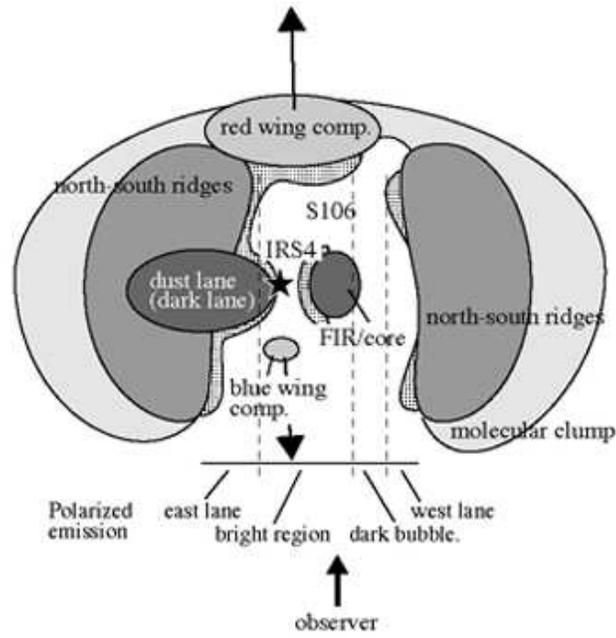}
\caption{
           Sketch of the relationship between the gas structure and polarization 
           intensity. The grayscale ellipses denote the characteristic gas structures. 
           The regions masked by hatched dots indicate the polarized 
           emitting regions. 
           }
\label{Figure 8}
\end{center}
\end{figure}

\clearpage

\begin{figure}
\begin{center}
\epsscale{0.30}
\plotone{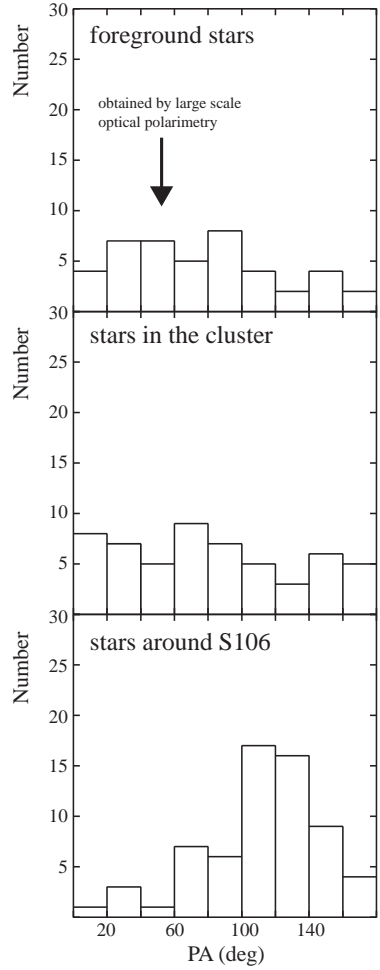}
\caption{
           Histograms of the polarization position angle of foreground stars (upper panel),
           stars in cluster S106 (middle panel), and stars around S106
           (lower panel). The arrow indicates the position angle of foreground stars
           obtained by Staude et al.\,(1982) using optical polarimetry.
           }
\label{Figure 9}
\end{center}
\end{figure}

\clearpage 

\begin{figure}
\begin{center}
\epsscale{0.50}
\plotone{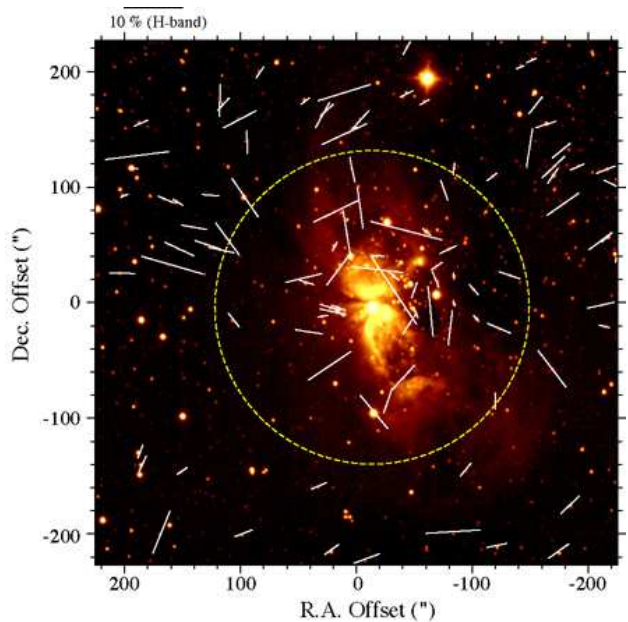}
\caption{
           Polarization vector of each point-like source on the $H$-band intensity image.
           The dotted circle denotes the cluster S106 area
           (see $\S$ 3.3).
           }
\label{Figure 10}
\end{center}
\end{figure}

\begin{figure}
\begin{center}
\epsscale{0.50}
\plotone{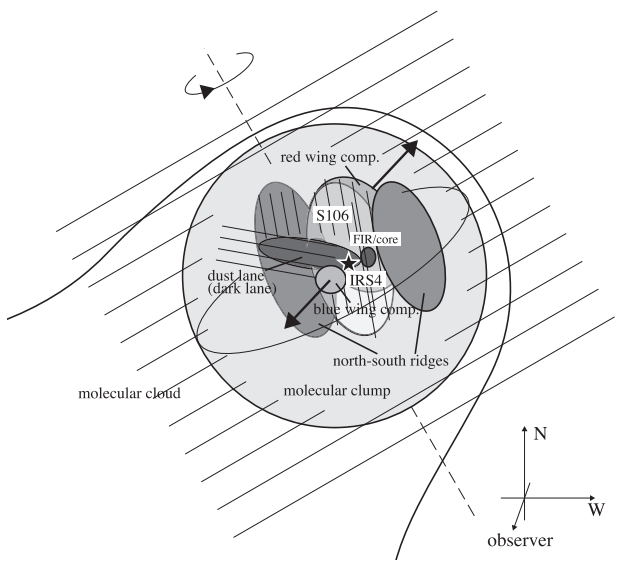}
\caption{
           Sketch of the relationship between the gas structure and the magnetic field
           structure. The grayscale ellipses denote the characteristic structures of the gas. 
           The thin lines show the magnetic field direction. The dashed line indicates
           the direction of axial rotation obtained by Lucas et al.\,(1978).
           }
\label{Figure 11}
\end{center}
\end{figure}









\clearpage

\end{document}